\documentclass[prl,twocolumn,groupedaddress,amsmath]{revtex4}
\usepackage{graphicx}
\usepackage{amsmath,amssymb}
\usepackage[utf8]{inputenc}

\newcommand{\rhoc}{\rho_{\scriptstyle\rm c}}

\begin{document}
\title{Random mobility and spatial structure often enhance cooperation}

\author{Estrella A. Sicardi}
\affiliation{Instituto de F\'{\i}sica, Facultad de Ciencias, Universidad
de la Rep\'ublica, Igu\'a 4225, 11400 Montevideo, Uruguay}
\affiliation{Instituto de F\'\i sica, Facultad de Ingenier\'\i a,
Universidad de la Rep\'ublica, Julio Herrera y Reissig 565, 1100
Montevideo, Uruguay}
\author{Hugo Fort}
\affiliation{Instituto de F\'{\i}sica, Facultad de Ciencias, Universidad
de la Rep\'ublica, Igu\'a 4225, 11400 Montevideo, Uruguay}
\author{Mendeli H. Vainstein}
\affiliation{Instituto de F\'\i sica, Universidade de Brasília,
CP 04513, 70919-97, Brasília, DF, Brazil}
\author{Jeferson J. Arenzon}
\affiliation{Instituto de F\'\i sica, Universidade Federal do
Rio Grande do Sul, CP 15051, 91501-970 Porto Alegre RS, Brazil}

\date{\today}

\begin{abstract}
The effects of an unconditional move rule in the spatial Prisoner's Dilemma,
Snowdrift and Stag Hunt games are
studied. Spatial structure by itself is known to modify the
outcome of many games when compared with a randomly mixed population, 
sometimes promoting, sometimes
inhibiting cooperation. Here we show that random dilution and
mobility may suppress
the inhibiting factors of the spatial structure in the Snowdrift
game, while enhancing the already larger cooperation found in
the Prisoner's dilemma and Stag Hunt games.

\noindent
{\bf Corresponding author}: Jeferson J. Arenzon, +55 51 33086446
(phone), +55 51 33087286 (fax), arenzon@if.ufrgs.br
\end{abstract}

\maketitle

\section{Introduction}

Competition and cooperation
are two inseparable sides of the same coin. While competition is
a key concept in Darwin's theory of evolution,
cooperation is rather puzzling, yet ubiquitous
in nature~\cite{SmSz97}.
How cooperative behavior evolved among self-interested individuals
is an important open question in biology and social sciences, along
with the issue of how cooperation contends with competition in order to achieve global 
and individual optima.
A powerful tool to analyze these problems
is evolutionary game theory~\cite{Smith82,Weibull95,HoSi98}, 
an application of the mathematical theory of games to
biological contexts.
Of particular relevance are two-player games where each player has a
strategy space containing two possible actions (2$\times$2 games),
to cooperate (C) or to defect (D). The payoff of a player depends on
its action and on the one of its co-player. Assuming pairwise, symmetric
interaction between the players, there are four possible values for 
this payoff. Cooperation involves a cost
to the provider and a benefit to the recipient. Two
cooperators thus get a {\em reward} $R$ while two defectors get
 a {\em punishment} $P$. The trade between
a cooperator and a defector gives the {\em temptation} $T$ for the
latter, while the former receives the {\em sucker's payoff}, $S$.
We renormalize all values such that $R=1$ and $P=0$.
The ranking of the above quantities defines the game they are
playing. The paradigmatic example is the {\em Prisoner's Dilemma} (PD) game
in which the four payoffs are ranked as $T>R>P>S$.
It clearly pays more to defect whatever the opponent's strategy:
 the gain will be $T>R$ if the other cooperates and $P>S$ if he defects.
The dilemma appears since if both play D they get $P$, what is worse
than the reward $R$ they would have obtained had they both played C.
The PD is related with two other {\em social dilemma}
games~\cite{Poundstone92,Liebrand83}.
In most animal contests (in particular those involving escalating conflicts), 
mutual defection is the worst possible outcome for both players,
and the damage exceeds the cost of being exploited, {\it i.e.}~ $T>R>S>P$. This game is called {\em chicken}~\cite{Rapoport66}
or {\em snowdrift} (SD).
On the other hand, when the reward surpasses the temptation to defect, {\it i.e.}~ $R>T>P>S$, the game becomes the
{\em Stag Hunt} (SH)~\cite{Skyrms04}.
The coordination of slime molds is an example of animal behavior that has 
been described as a stag hunt~\cite{StZhQu00}. When individual amoebae of
{\it Dictyostelium discoideum} are starving, they aggregate to form one
large body whose reproductive success depends on the cooperation of many
individuals. Here,  we consider for $T>1$ the PD ($S<0$) and the SD
($S>0$), whose interface ($S=0$) is known as the weak form of the PD 
game, with $S=P=0$.  The Stag Hunt (SH) game is obtained for $S<0$ and $T<1$.

Classical evolutionary game theory
constitutes a mean-field-like approximation which
does not include the effect of spatial, correlated structures of populations.
Axelrod~\cite{Axelrod84} suggested to place the agents
on a two-dimensional spatial array interacting with their
neighbors. This cellular automaton was explored by Nowak and
May~\cite{NoMa92}, who 
found that such spatial structure allows cooperators to build
clusters in which the benefits of mutual cooperation can outweigh
losses against defectors, thus enabling cooperation to be sustained,
in contrast with the spatially unstructured game, where
defection is always favored. The original Nowak-May model was
extended and modified in several different ways 
(see ref.~\cite{SzFa07} and references therein).
Related to the present work, the effects of dilution and mobility 
were recently studied in refs.~\cite{VaAr01,VaSiAr07} 
in the weak form of the PD game.

Mobility effects are difficult to anticipate. When non-assortative movements
are included, the effective number of neighbors increases (towards the random mixing 
limit). Moreover, those clusters so necessary to sustain cooperation may 
now evaporate. Both these effects promote defection and the number of cooperators
is expected to decrease. On the other hand, dilution~\cite{AlTa08}
and mobility decrease the 
competition for local resources and help to avoid retaliation and abuse 
(although that may require more contingent movements), thus
tending to increase cooperation. In the evolutionary game context, diffusion was studied by 
several authors, sometimes as a cost to wander between patches without spatial 
structure~\cite{DuWi91,EnLe93} or as a trait connected with laying offspring 
within a given dispersal range~\cite{BaRa98,Koella00,HaTa05,LeFeDi05}.
An explicit diffusive process was studied in the framework of the
replicator equation~\citep{HoSi98}, extended to include a diffusive 
term~\citep{FeMi95,FeMi96}; however, the interactions were still 
mean field like.
Aktipis~\citep{Aktipis04} considered contingent movement of
cooperators with a ``win-stay, lose-move'' rule, allowing them to
invade a population of defectors and resist further invasions.
Models with alternating viscosities, which reflect different stages of
development that can benefit from the clusterization of cooperators
or from dispersal, have also been considered and promote altruism,
since the high viscosity phase allows interactions between close relatives
and the low viscosity phase reduces the disadvantages of local competition
among related individuals. In cases of populations with only a highly viscous
phase, the effects of interactions  among relatives and competition for
local resources tend to balance and thus the evolution of altruistic behavior
is inhibited~\cite{WiPoDu92,Taylor92}. Differently from previous
works, ref.~\cite{VaSiAr07} considered a diluted version of
Nowak-May's spatial PD model where individuals are able to perform
random walks on the lattice when there is enough free space (the
non-assortative ``always-move''
strategy). Specifically,
the setting was the simplest possible: random
diffusion of simple, memoryless, unconditional, non-retaliating, strategy-pure
 agents. Under these conditions, cooperation was found not only to be 
possible and robust but, for a broad range of the parameters (density, viscosity, etc),
often enhanced when compared to the strongly viscous (no
mobility) case. The
parameters chosen put the model at the interface between the PD and the SD, and
a natural question is how robust is the behavior when $S<P$, that is, in  a
genuine PD game? Moreover, how does mobility affect other games, like the SD or the SH?
Recently, Jian-Yue {\em et al}~\cite{JiZhYi07} extended the results of
ref.~\cite{VaSiAr07} for the SD game and COD dynamics (see next section), 
but with a restricted choice of $S$ and $T$. Another relevant question
regards the existence of any fundamental difference between those games 
when mobility is introduced.
In particular, in those cases where the spatial structure is known to
inhibit cooperation~\cite{HaDo04}, does mobility change this picture?

 Our objective here is to present a more comprehensive analysis and extend 
our previous study in several directions, trying to shed some light on the 
above questions. The paper is organized as follows. The following section 
describes the details of the model and simulation. Then, we present the 
results for two possible implementations depending on the order of the 
diffusive and offspring steps. Finally, we present our conclusions and 
discuss some implications of the results.

\section{The Model}
\label{section.model}

The model is a two dimensional stochastic cellular automaton in which cells
are either occupied by unconditional players (cooperators or
not) or vacant. At time $t$, the variable $S_i(t)$ is 0 if the corresponding
lattice cell is empty, or $\pm 1$ depending on 
whether the agent at that site cooperates (1) or defects ($-1$).
The relevant
quantity is the normalized fraction of cooperators, $\rhoc$, after
the stationary state is attained, defined as
$\rhoc=(1+M/\rho)/2$, where $M$ is the ``magnetization'' 
$M=N^{-1}\sum_i\langle S_i(\infty)\rangle$,
$N$ is the system size and $\rho\neq 0$ is the fraction of
occupied sites, that is kept fixed at all times (when $\rho \neq  1$, 
we call the system diluted). The symbol 
$\langle\ldots\rangle$ stands for an
average over the ensemble of initial configurations. We
call ``active'' a site that has changed strategy since the previous
step. 
At each time step, all individuals play against all its four 
nearest neighbors (if present), collecting the payoff from the
combats. After that,
it may either move or try to generate its offspring.
We consider a best-takes-all reproduction: each player compares its 
total payoff
with those of its nearest neighbors and changes strategy following
the one (including itself) with the greatest payoff among them.
This strategy changing updating rule preserves the total
number of individuals, thus keeping $\rho$ constant. If a tie among
some of the neighbors occurs, one of them is randomly chosen. 
During the diffusive part,
each agent makes an attempt to jump to a nearest neighbor site chosen randomly,
that is accepted, provided
the site is empty, with a probability given by the mobility
parameter $m$. Notice that $m$ does not measure the effective mobility,
that depends on both $m$ and $\rho$, but only a tendency to move, space
allowing. However, different combinations of both parameters could give
the same effective mobility (measured, for example, through the mean
square displacement), but that parameter alone would not suffice to
characterize the game, since spatial correlations are determined by $\rho$.
Among the several ways of implementing the reproductive and
diffusive steps, here we consider
two possibilities, named contest-offspring-diffusion (COD) and
contest-diffusion-offspring (CDO). In the former, as the name says, 
each step consists of combats
followed by the generation of offspring done in parallel,
and then diffusion, while in the latter, the diffusion and
offspring steps are reversed. Of course, the stochasticity introduced by the 
mobility disappears under some conditions (e.g., $\rho=1$ or
$m=0$), and the final outcome may now depend on the initial
conditions.

We explore many choices of the payoff parameters $T$ and $S$ while
$P$ and $R$ are kept fixed at 0 and 1, respectively, and compare
the effects of diffusion with those obtained either for the related
spatial weak version~\cite{VaSiAr07} or for the randomly mixed limit.
Square lattices of sizes ranging from $100 \times 100$ to
$500 \times 500$ with periodic boundary conditions are used.
Averages are performed over at least 100 samples and equilibrium is usually 
attained before 1000 network sweeps are completed, although in some
cases not even $10^7$ sweeps are enough to bring the system to an
equilibrium fixed point. The initial 
configuration has $\rho N$ individuals placed randomly on the lattice,
equally distributed among cooperators and defectors.

\section{Results}

\begin{figure}[htb]
\includegraphics[width=7.8cm]{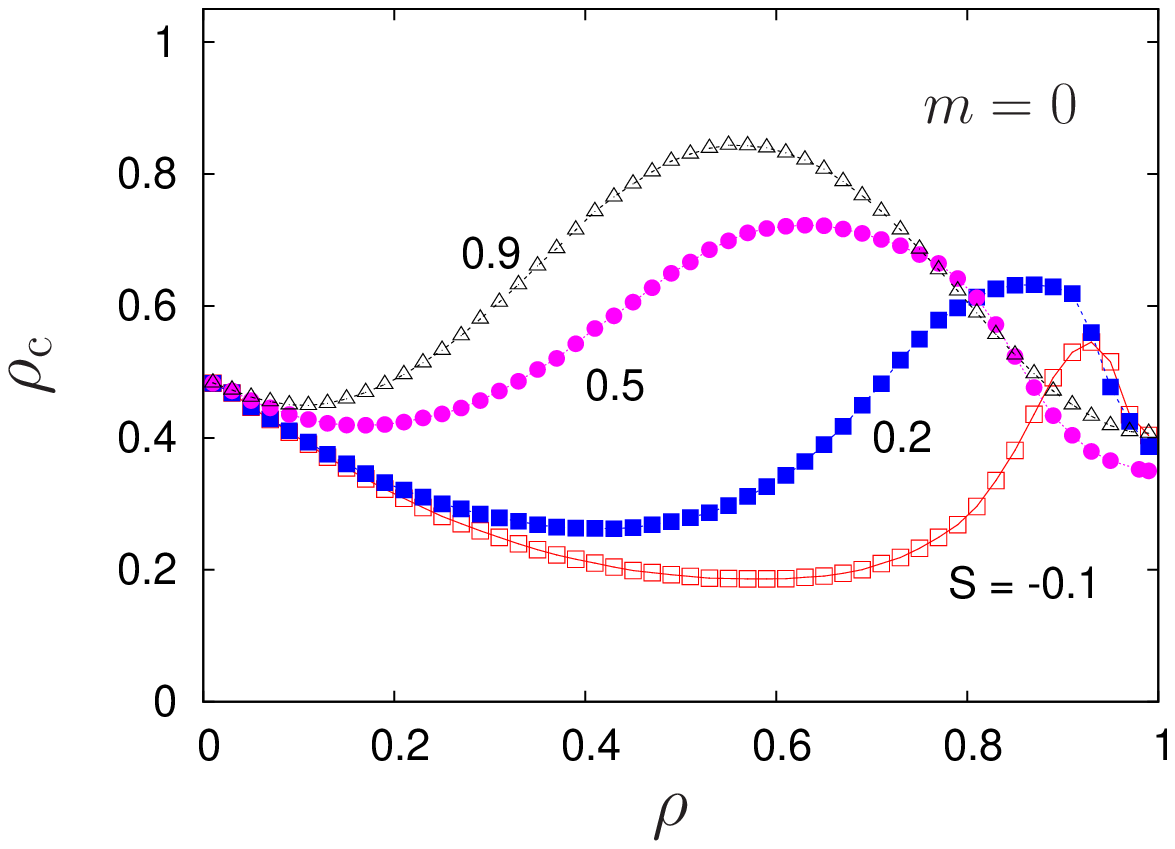}
\includegraphics[width=7.8cm]{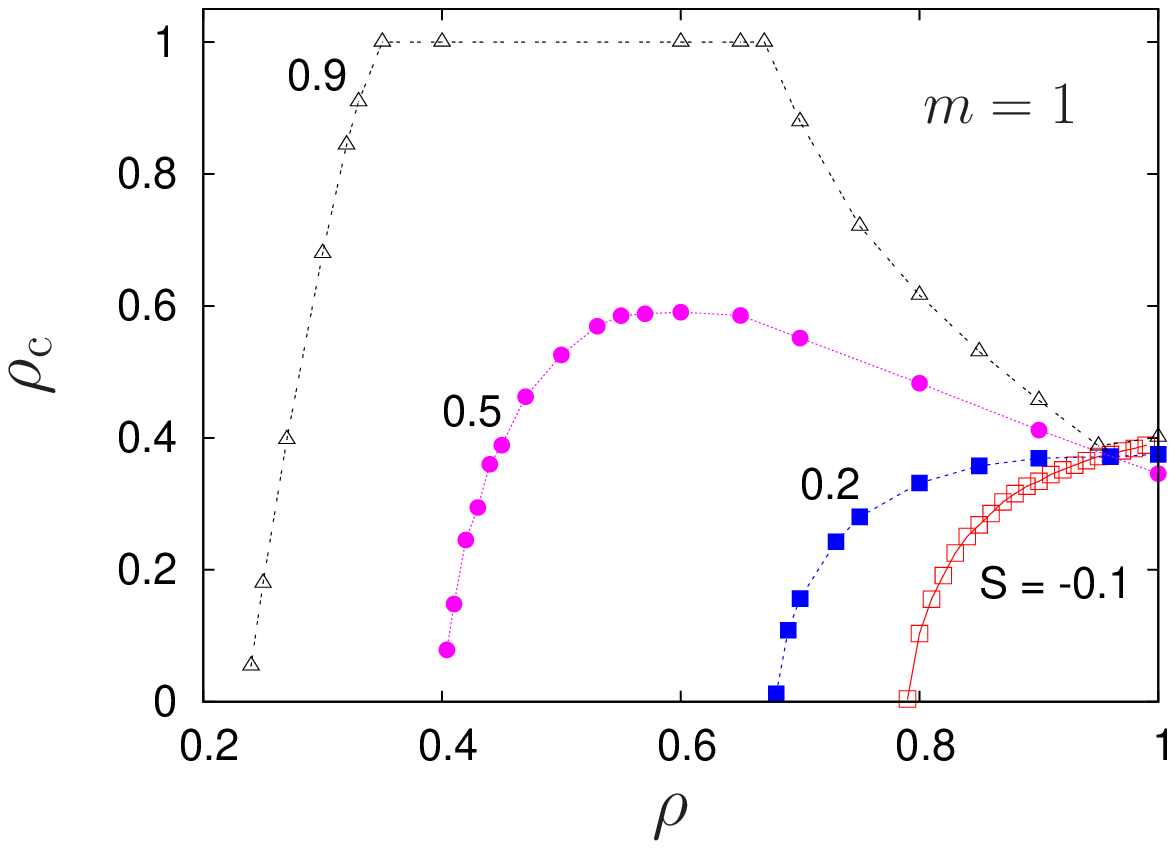}
\caption{Average fraction of cooperating individuals $\rhoc$  
for mobilities $m=0$ (top) and 1 (bottom), 
$P=0$, $R=1$, $T=1.4$ and several values
of $S$ in the COD case. The case $S=-0.1$ is representative of the whole
interval around the weak PD point $S=0$ ($-0.2<S<2/15$). Indeed, all
curves in this interval collapse. Notice also that near $\rho=1$, 
negative responses occur and $\rhoc$ may decrease as $S$ gets larger,
the effect being stronger for $m=0$ (see text). At small densities,
mobility is detrimental to cooperators: isolated clusters of cooperators, 
that could survive for $m=0$, can be predated by mobile defectors once 
mobility is considered. At intermediate densities, mobility can
strongly increase the amount of cooperation.}
\label{fig.COD_rhoc2}
\end{figure}

\begin{figure*}[th]
\includegraphics[width=8cm]{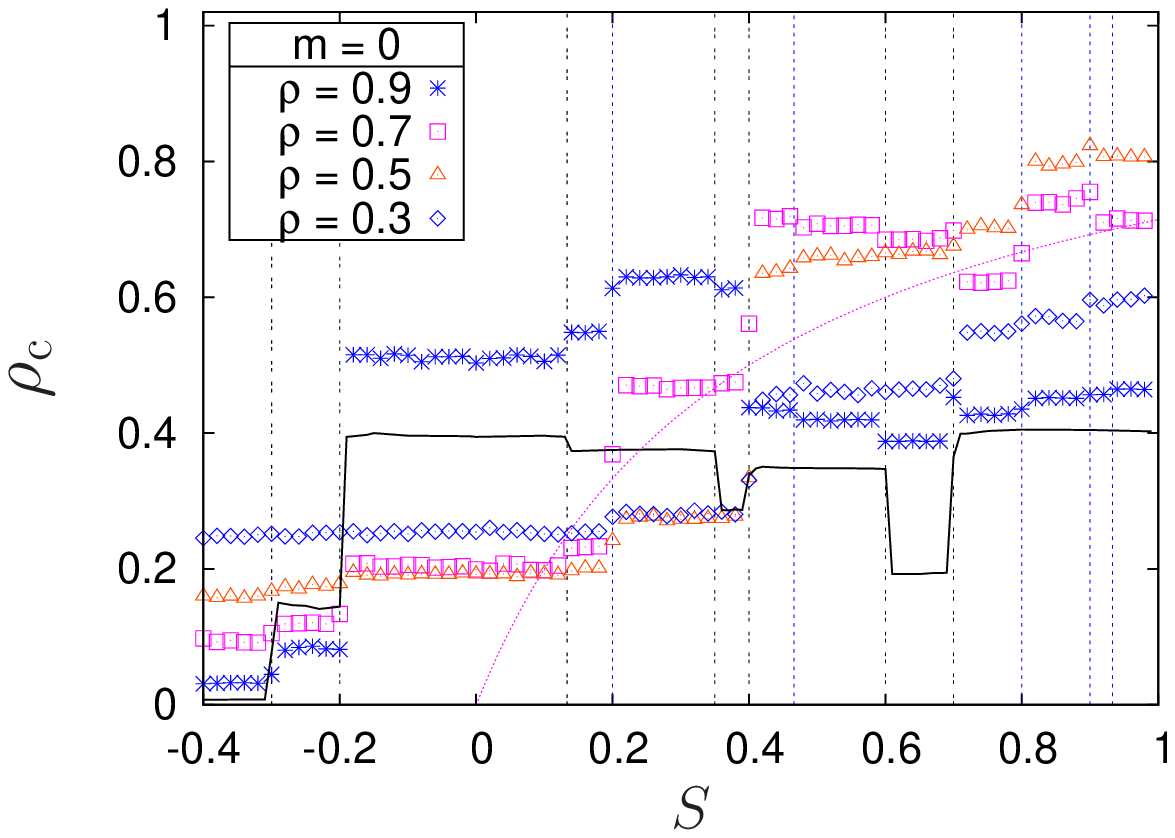}
\includegraphics[width=8cm]{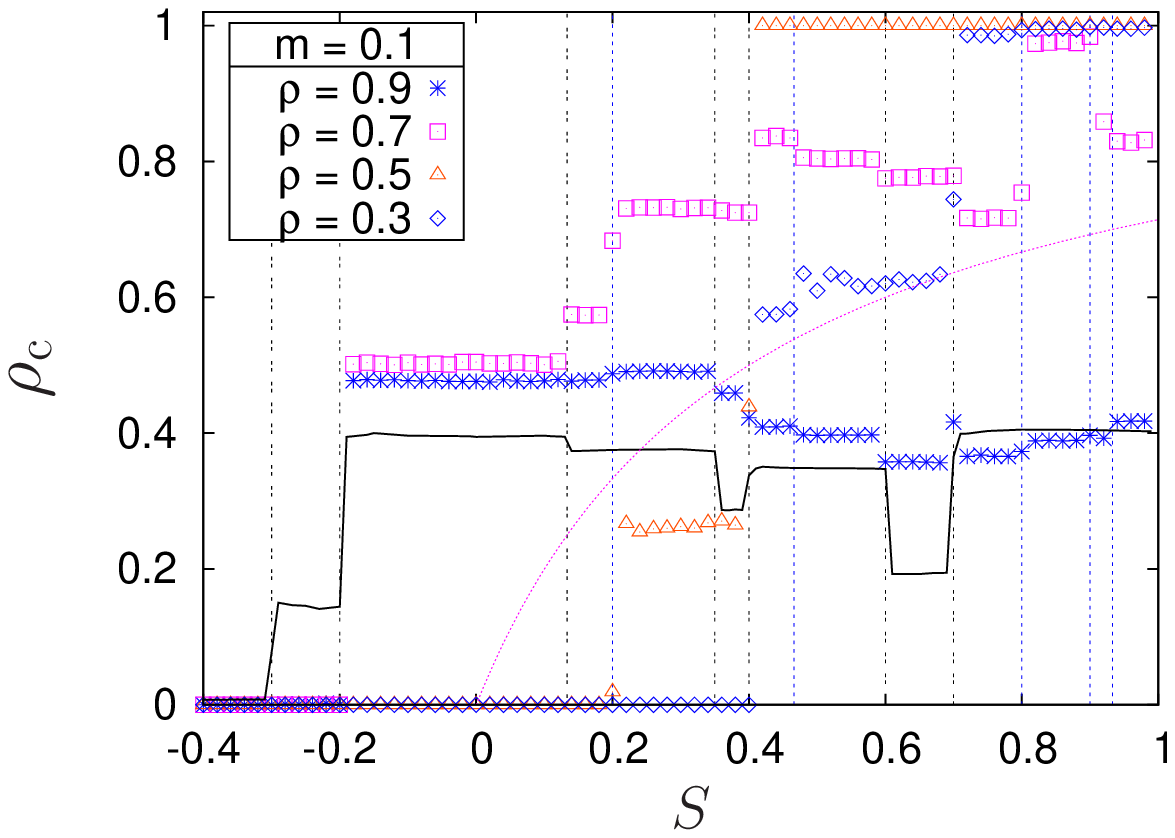}
\includegraphics[width=8cm]{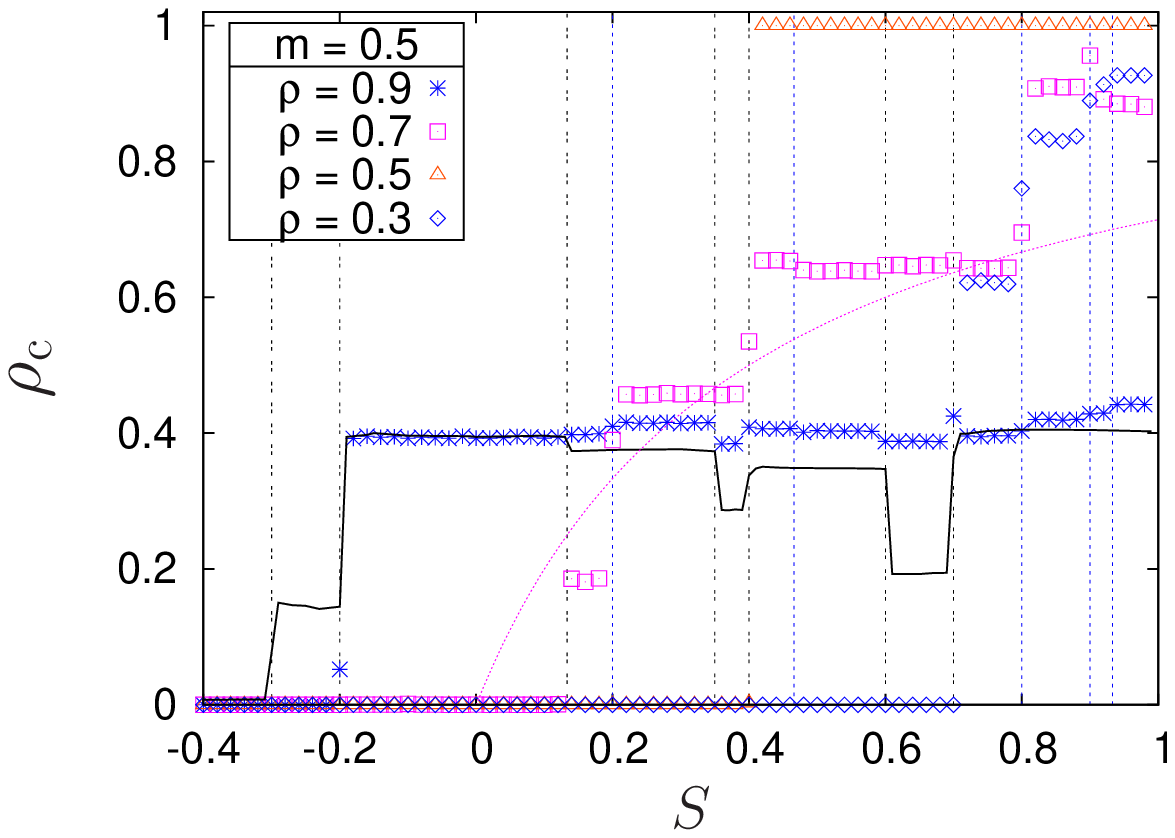}
\includegraphics[width=8cm]{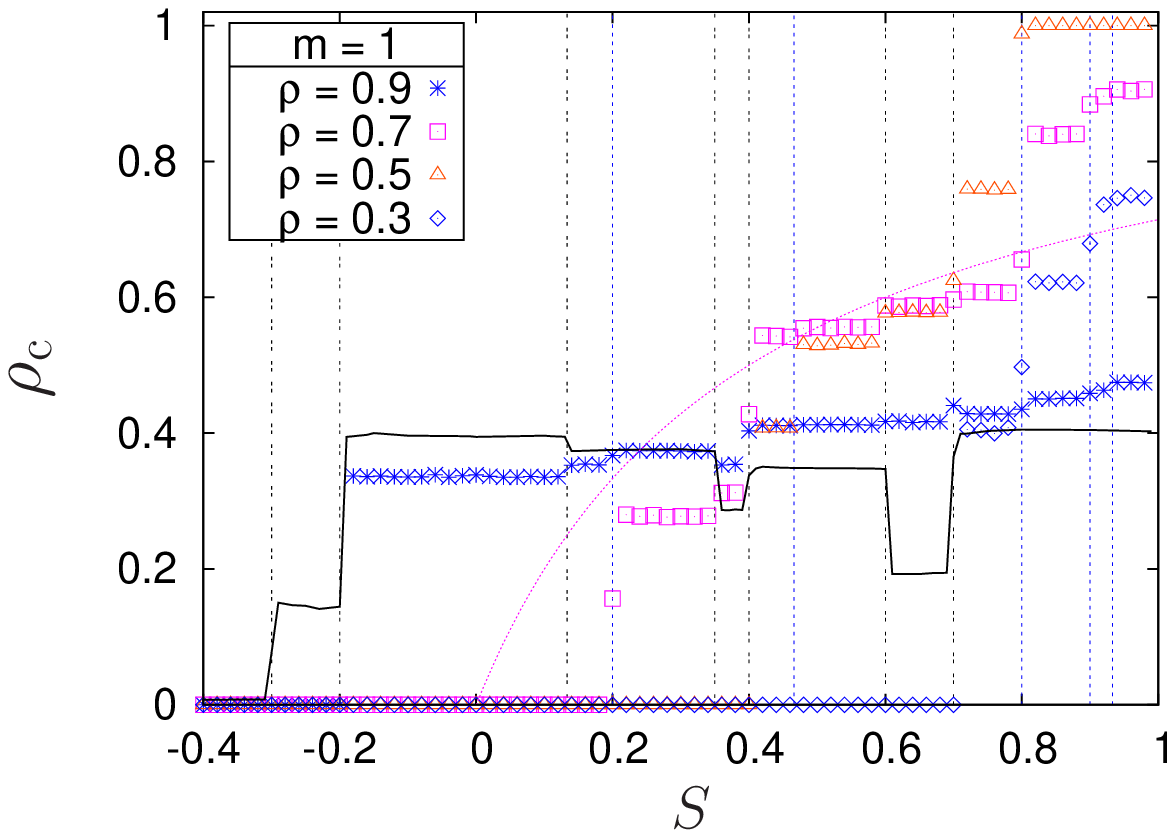}
\caption{Average fraction of cooperating individuals $\rhoc$ versus
  $S$ ($T=1.4$, $R=1$ and $P=0$) in the SD ($S>0$) and PD ($S<0$)
  games with COD dynamics for different values of the mobility and
  $\rho$. The solid black line is the case $\rho=1$. The vertical dashed lines 
  locate the transition points for $\rho=1$ ($S=-0.3$,
  $-0.2$, $2/15$, $0.35$, $0.4$, $0.6$ and $0.7$) and the vertical dotted lines locate a few more when $\rho<1$
  ($S=0.2$, $7/15$, $0.8$, $0.9$ and $14/15$), as explained in the text. The
  curved line is the expected result for random
  mixing ($\rhoc=0$ for the PD game, $S<0$ and eq.~\ref{eq.mf} for the
  SD game, $S>0$).
  The point $S=0.6$ corresponds to the standard parametrization
  $T=1+r=1.4$, $R=1$, $S=1-r=0.6$ and $P=0$ for the SD game,
  considered in \cite{JiZhYi07}. For comparison, the case without
  mobility ($m=0$), whose results are obviously independent of the
  diffusion dynamics, is also shown. No new transitions appear due to
  mobility besides those already present when $m=0$ for the COD 
  dynamics~\cite{VaAr08}. A further non trivial effect is also noticeable: for some fixed
  values of $S$ and $\rho$ (e.g., $S=0.5$ and $\rho=0.7$), there is an 
  optimal value of $m$ where $\rhoc$ is maximized.}
\label{fig.COD_S}
\end{figure*}

\begin{figure}[htb]
\includegraphics[width=7.8cm]{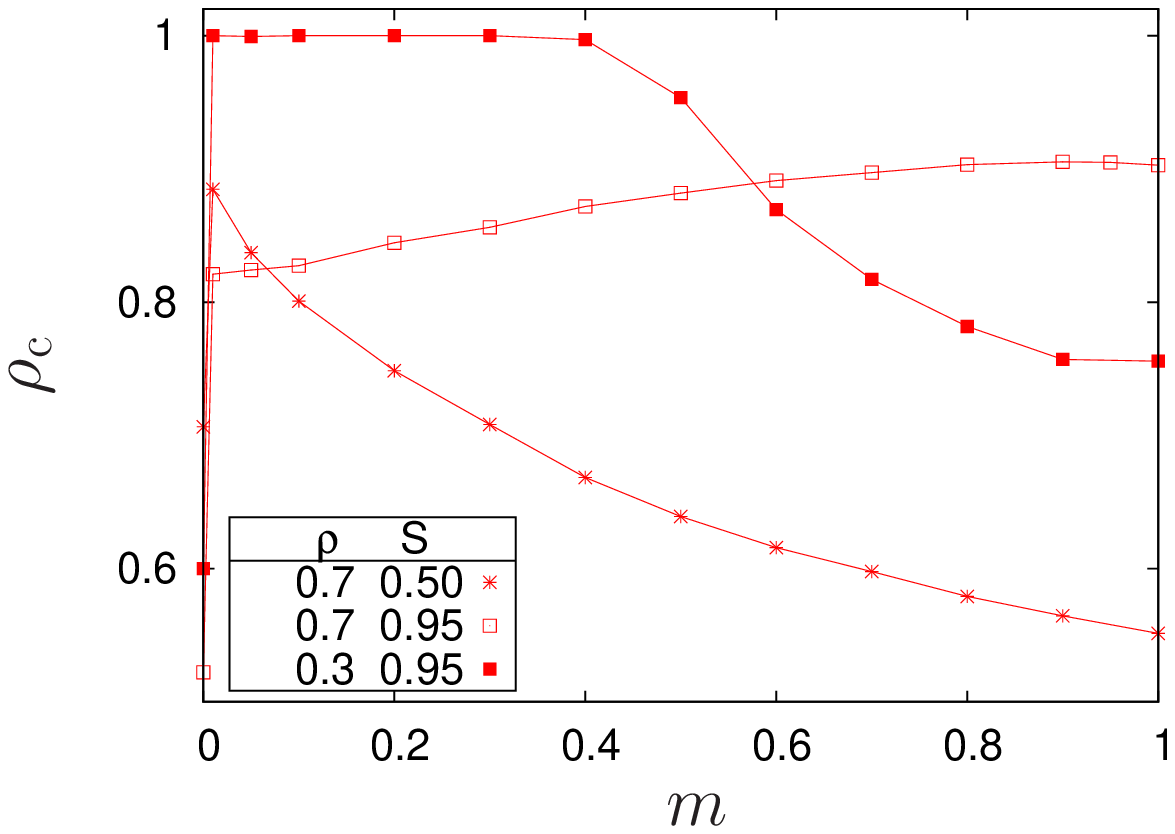}
\caption{Average fraction of cooperating agents $\rhoc$ as a function of
the mobility $m$ for several values of $\rho$ and $S$. Notice that $\rhoc$ may
either increase ($m=1$ is optimal) or decrease (a small, but non zero, $m$ is
optimal), depending on the parameters, and that there is
a discontinuity at $m=0$.}
\label{fig.COD_m}
\end{figure}

Ref.~\cite{VaSiAr07} considered only  the weak limit  $P=S=0$ of the PD
game for $T=1.4$ and $R=1$, that is, inside the region where, for $\rho=1$, 
both strategies coexist along with active sites. When $S\neq 0$,
this region may comprise values of $S$ both in the PD and in the SD 
regimes, that is, $S<0$ and $S>0$, respectively.
Figs.~\ref{fig.COD_rhoc2}
and \ref{fig.COD_S} show that  the results found in ref.~\cite{VaSiAr07}
with the COD dynamics remain unchanged for this rather broad range of $S$.
Indeed, the line labeled $S=-0.1$ in fig.~\ref{fig.COD_rhoc2} exactly
superposes with the results for all values of $S$ in the aforementioned
interval, $S=0$ included. This can be more clearly seen in
fig.~\ref{fig.COD_S}, where $\rho$, $m$ and $S$ are varied, and
a plateau in the interval $(-0.2,2/15)$~\cite{Hauert01} is observed 
(although the value of $\rhoc$ in the plateau depends on both $\rho$ and
$m$). Besides this active phase, the system presents a large number 
of different phases, with sharp transitions between them.
For comparison, the no mobility case~\cite{VaAr08}, 
$m=0$, is also included, as well as the cooperators 
density when $\rho=1$ (solid black line). The
transition points, calculated considering the several possible local 
neighborhoods~\cite{VaAr08},  are represented by vertical dashed and 
dotted lines located at the points where $S$ equals to
$(T+3P)/2-R=-0.3$, $2T+2P-3R=-0.2$,
$(T+3P-R)/3=2/15$, $(T+3P)/4=0.35$, $T+P-R=0.4$, 
$(2P+2T-R)/3=0.6$ and $P+T/2=0.7$.
In addition to these transitions, a few more are introduced when the system 
is diluted (vertical dotted lines), $\rho<1$, each phase being characterized
by the fraction of cooperators and by the way they organize 
spatially~\cite{VaAr08}. 
When mobility is introduced after the offspring generation (COD dynamics), 
no new transition appears. Dilution (without mobility)
allows cooperation for small values of $S$ ($S<-0.3$),  which is absent with
$\rho=1$. 
For  low densities in particular, clusters are
small and isolated; therefore, depending on the initial condition, cooperators
succeed in forming pure clusters. 
However, as soon as mobility is introduced, this disorder driven
phase disappears since small cooperative clusters are easily predated by 
wandering defectors.
In this low density situation, cooperation
is only sustained when the exploitation is not too strong (larger values of
$S$), as can be seen in the case $\rho=0.3$ and $m=1$, where cooperation
exists only for $S>0.7$. For intermediate densities, some
phases may coalesce, like the large $S$ region for $\rho=0.5$ where $\rhoc=1$.
Interestingly, mobility has a non trivial effect in the negative
response that is already present at $\rho=1$ or $m=0$. When the sucker's 
payoff $S$ increases (less exploitation), one expects higher levels of
cooperation. But the opposite behavior is sometimes observed, and the
number of cooperators may also decrease. An example occurs when $\rho=1$ as $S$ 
increases beyond the transition point 0.6, and $\rhoc$ attains a new plateau, far
below the previous one. However, with mobility such effect may be
enhanced, attenuated or reverted. Fig.~\ref{fig.COD_S} depicts many
examples of such behavior.
Moreover, for a fixed $\rho$, mobility affects different phases in 
diverse ways. For example, for $\rho=0.5$, when $m$ changes from 0.1
to 0.5, the cooperative phase at $0.2<S<0.4$ disappears, while the one for $0.4<S<1$
suffers no alteration. When $m=1$, instead, this region splits into 
five smaller regions. Therefore, whether large or small mobility is better for
cooperators strongly depends on both $m$ and $\rho$. For example, for
$\rho=0.9$, $\rhoc$ increases with $m$ when $S$ is large and decreases
for smaller values. The behavior of $\rhoc$ as a function of the mobility
is shown in fig.~\ref{fig.COD_m}. In some cases cooperation is an increasing
function of the mobility and the optimal value is thus $m=1$. On the
other hand, it may also be detrimental for cooperation, and $\rhoc$ steadily
decreases with $m$. In this latter case, a non zero but very small mobility
gives the optimal value.

The most important result is obtained when
we compare the simulations with what one would obtain in a large randomly 
mixed population. A simple mean field~\cite{HoSi98,Hauert01} argument
leads to three possible solutions:  two absorbing states, $\rhoc=0$ and
1, and a mixed case with
\begin{equation}
\rhoc=\frac{S}{S+T-1},
\label{eq.mf}
\end{equation}
where we have already considered $R=1$ and $P=0$. Depending on the
values of $S$ and $T$, one of these solutions may become the stable
one. In the mean-field PD case, there is no cooperation and $\rhoc=0$ is
the stable solution. For the SH, the stable solution depends also
on the initial density of cooperators $\rhoc^0$ and eq.~\ref{eq.mf}
delimits the basin of attraction of each solution: $\rhoc=0$ 
if $\rhoc^0<S/(S+T-1)$ and 1 otherwise. The solution given by eq.~\ref{eq.mf}, 
shown in fig.~\ref{fig.COD_S} as a curved line, 
is only stable in the SD game. It has been known that a structured spatial
distribution of agents often inhibits cooperation~\cite{HaDo04} in the
SD game, as opposed to what happens in the PD and SH games. 
Nevertheless, when both dilution and mobility are introduced, 
cooperation is not so often inhibited.  
 Indeed, in fig.~\ref{fig.COD_S} one can see that in many cases the
 spatially 
distributed population outperforms the randomly mixed population in terms of cooperative behavior.

\begin{figure}[htb]
\includegraphics[width=7.8cm]{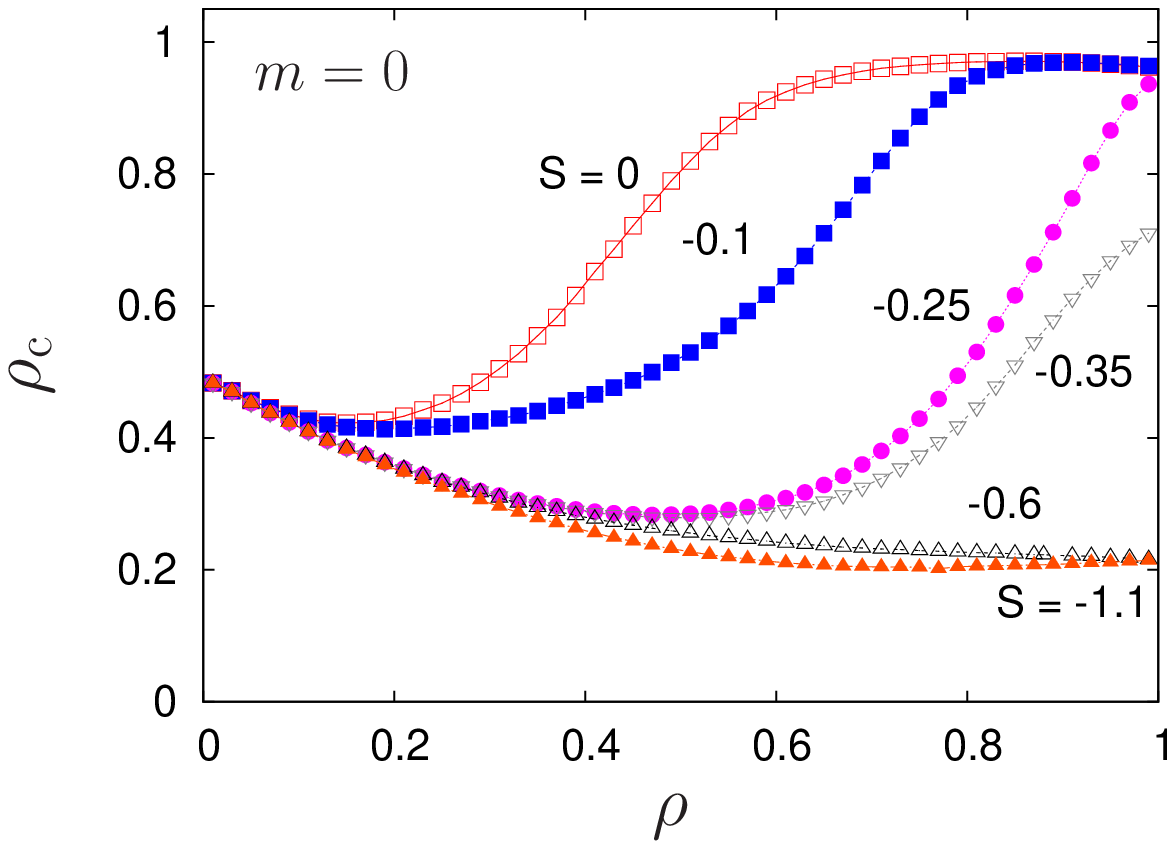}
\includegraphics[width=7.8cm]{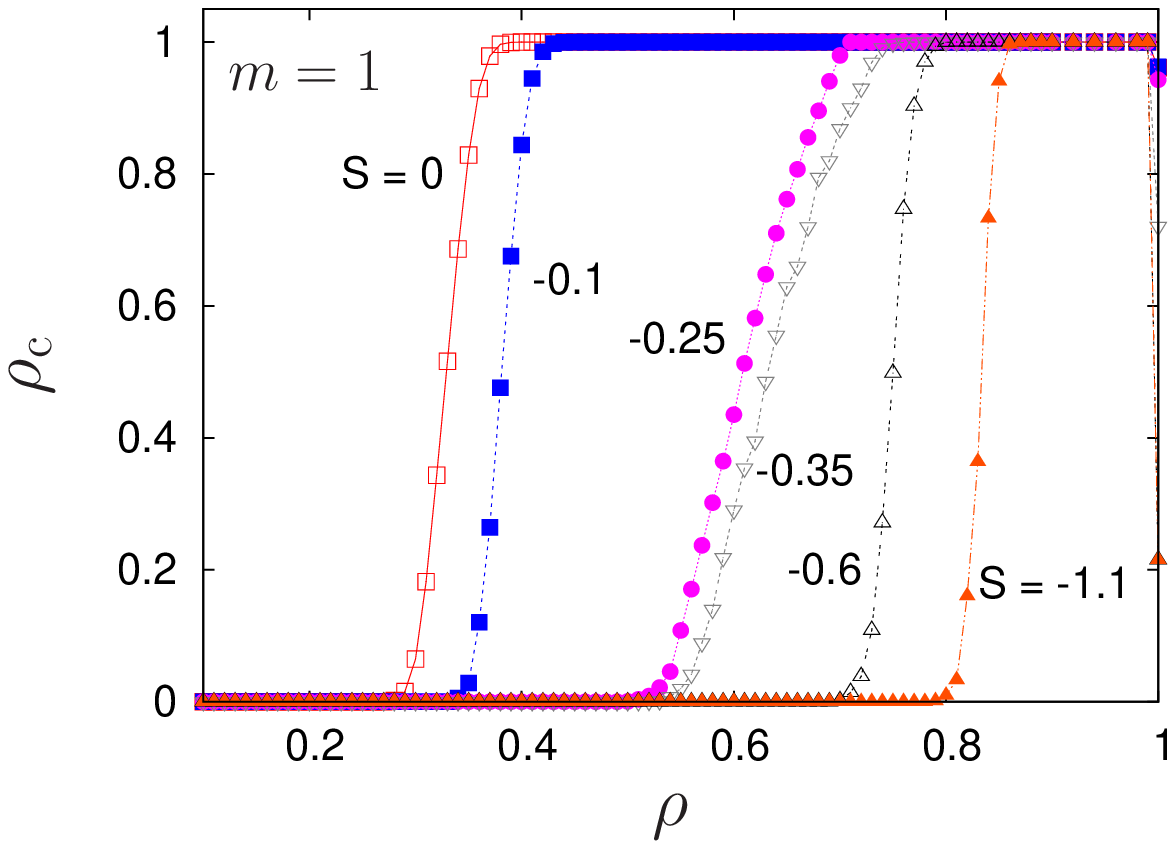}
\caption{Average fraction of cooperating individuals $\rhoc$  
for mobilities $m=0$ (top) and $m=1$ (bottom), $P=0$, 
$R=1$, $T=0.9$ and several values
of $S$ with COD dynamics for the SH game. Notice the strong
effect of the inclusion of mobility. The chosen values correspond to
the different plateaux seen in fig.~\ref{fig.COD_S_SH}. 
 The density of cooperators is a monotonically
increasing function of the total density for $m\neq 0$, while it is either
non-monotonic or monotonically decreasing for $m=0$.}
\label{fig.sh_rhocm1}
\end{figure}

\begin{figure*}[th]
\includegraphics[width=8cm]{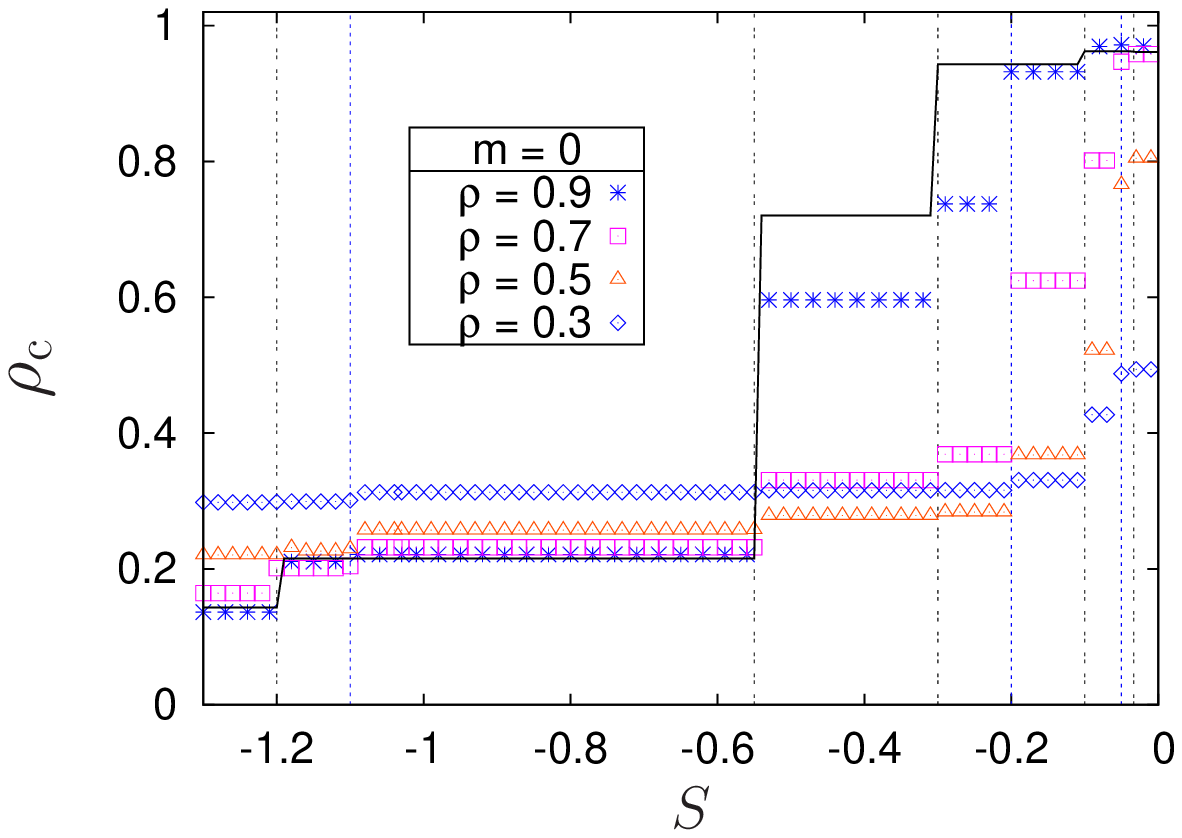}
\includegraphics[width=8cm]{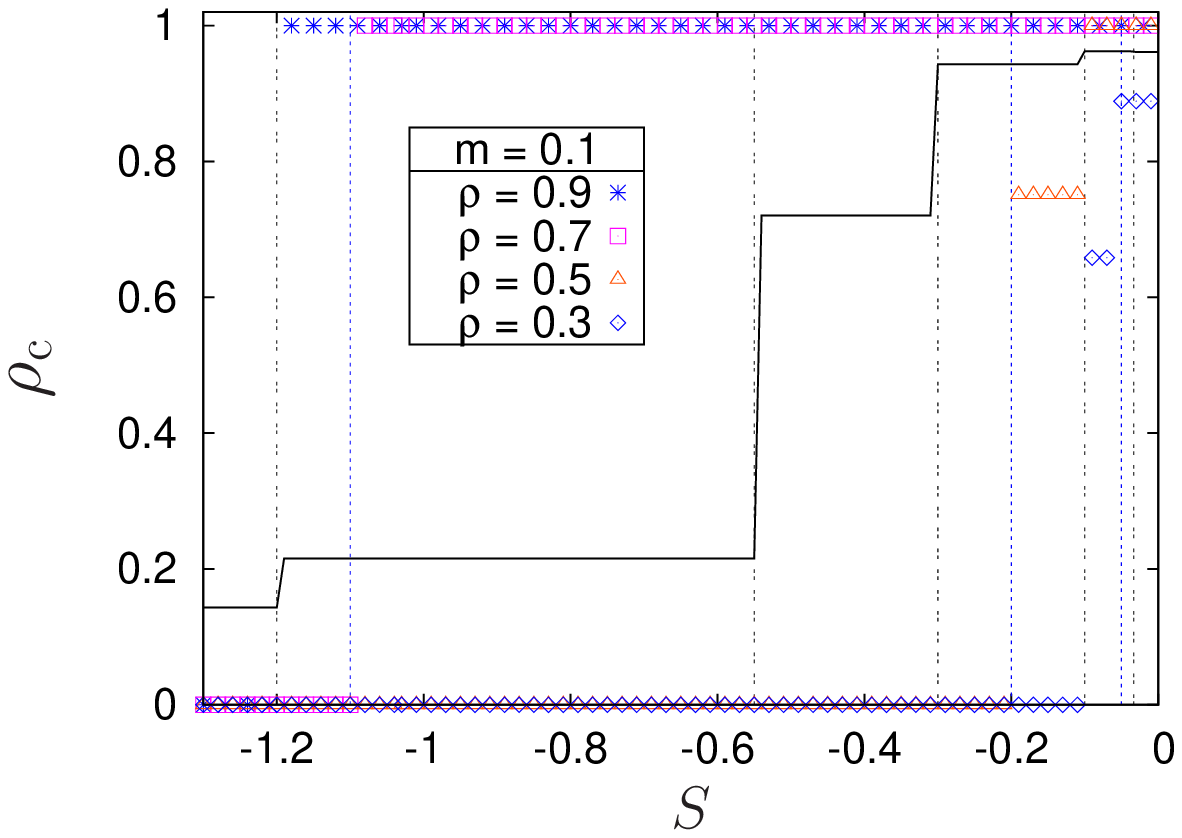}
\includegraphics[width=8cm]{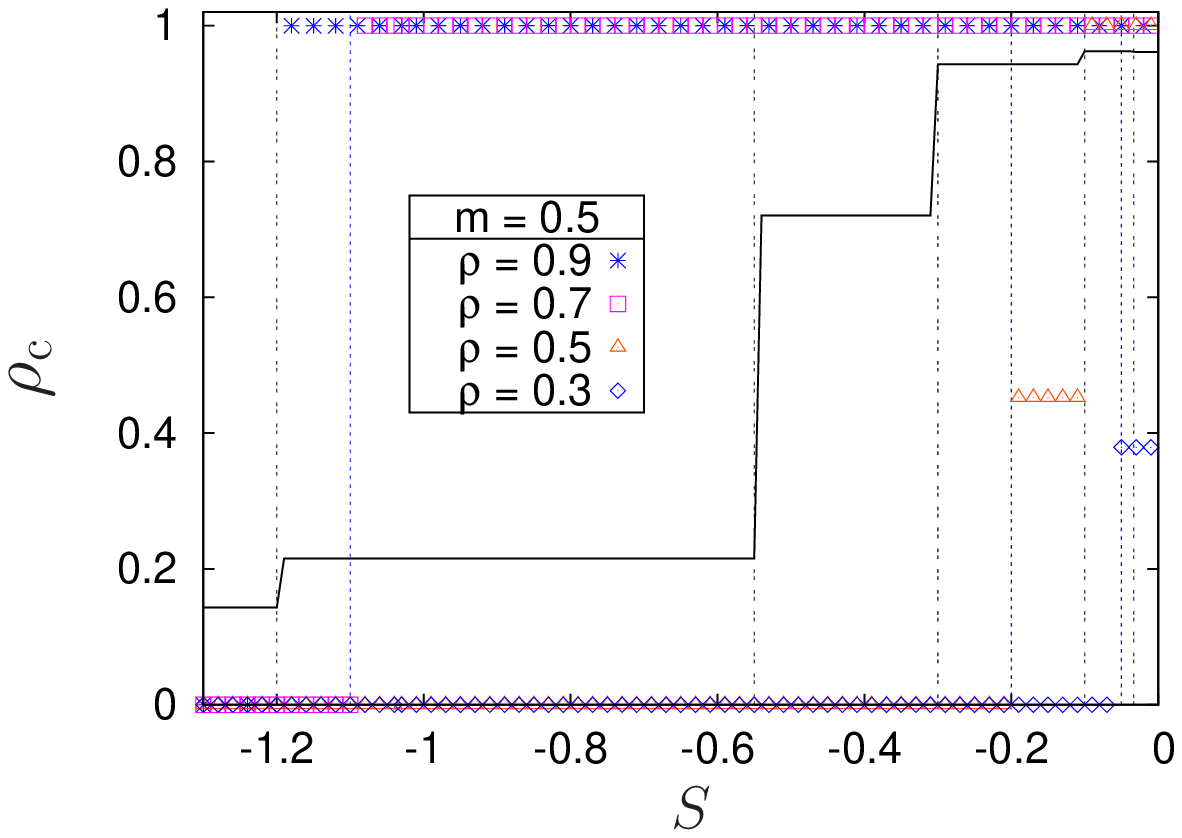}
\includegraphics[width=8cm]{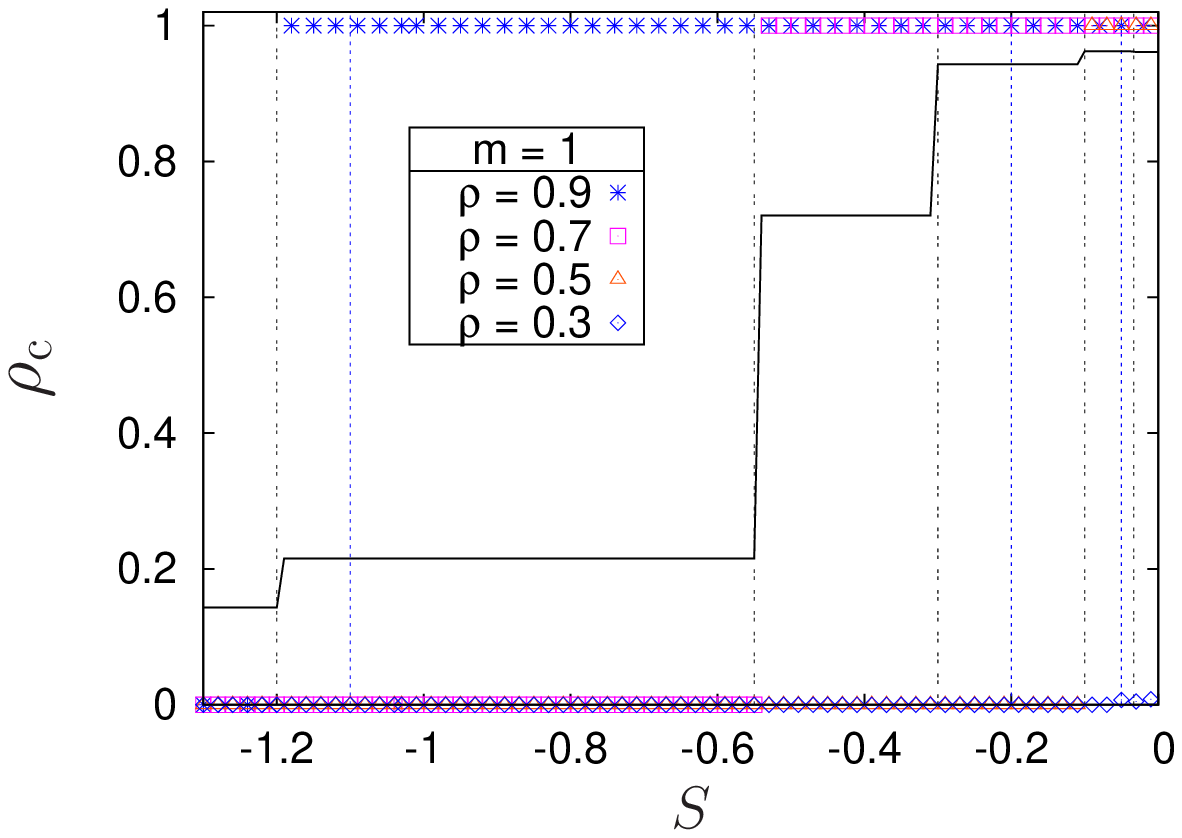}
\caption{Average fraction of cooperating individuals $\rhoc$  
versus $S$ ($T=0.9$, $R=1$ and $P=0$) for different values 
of the mobility and $\rho$ in the COD case. The solid black line is the
case $\rho=1$, while the vertical dashed lines locate the
transition points for $\rho=1$ ($S=-1.2$, $-0.55$, $-0.3$, $-0.1$ and $-1/30$) and the vertical dotted lines locate a few more when 
$\rho<1$ ($S=-1.1$, $-0.2$ and $-0.05$), as explained in the text. 
For comparison, the case without
mobility ($m=0$), whose results are obviously independent of the diffusion
dynamics, is also shown. No new transition appears due to mobility for the COD dynamics besides
those already present when $m=0$.}
\label{fig.COD_S_SH}
\end{figure*}

Besides the PD and SD games, we also studied the effect of mobility in
the SH game. Fig.~\ref{fig.sh_rhocm1} shows the normalized fraction
of cooperators as a function of the total density for several values of
$S<0$ and $T=0.9 < R=1$ when the mobility is either high ($m=1$) or
absent ($m=0$). Without mobility, isolated clusters of cooperators are
able to survive and even at very low densities cooperation is sustained.
Once mobility is included, cooperation at low densities is destroyed as
small cooperator clusters are easily predated by mobile defectors.
On the other hand, at higher densities cooperation is strongly enhanced
and $\rhoc=1$ for all values of $S<0$. It is interesting to notice
that mobility also changes the dependence on the density: with $m=1$
all curves are monotonically increasing functions of $\rho$, while
for $m=0$ they are always decreasing for low $S$ and non-monotonic
for larger values. The general dependence on $S$,
for several values of the mobility can be observed in fig.~\ref{fig.COD_S_SH}.
No negative response appears and the amount of cooperation increases with $S$,
as expected. 

\begin{figure}[th]
\includegraphics[width=7.8cm]{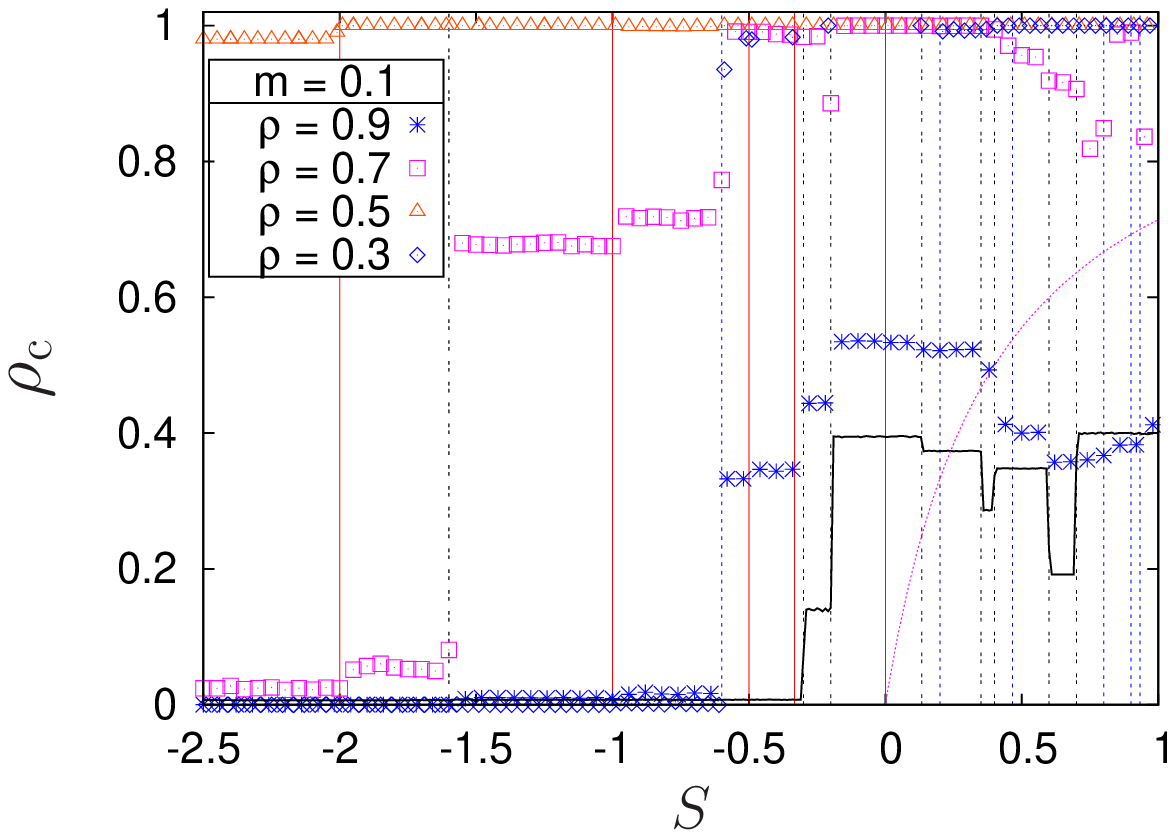}
\includegraphics[width=7.8cm]{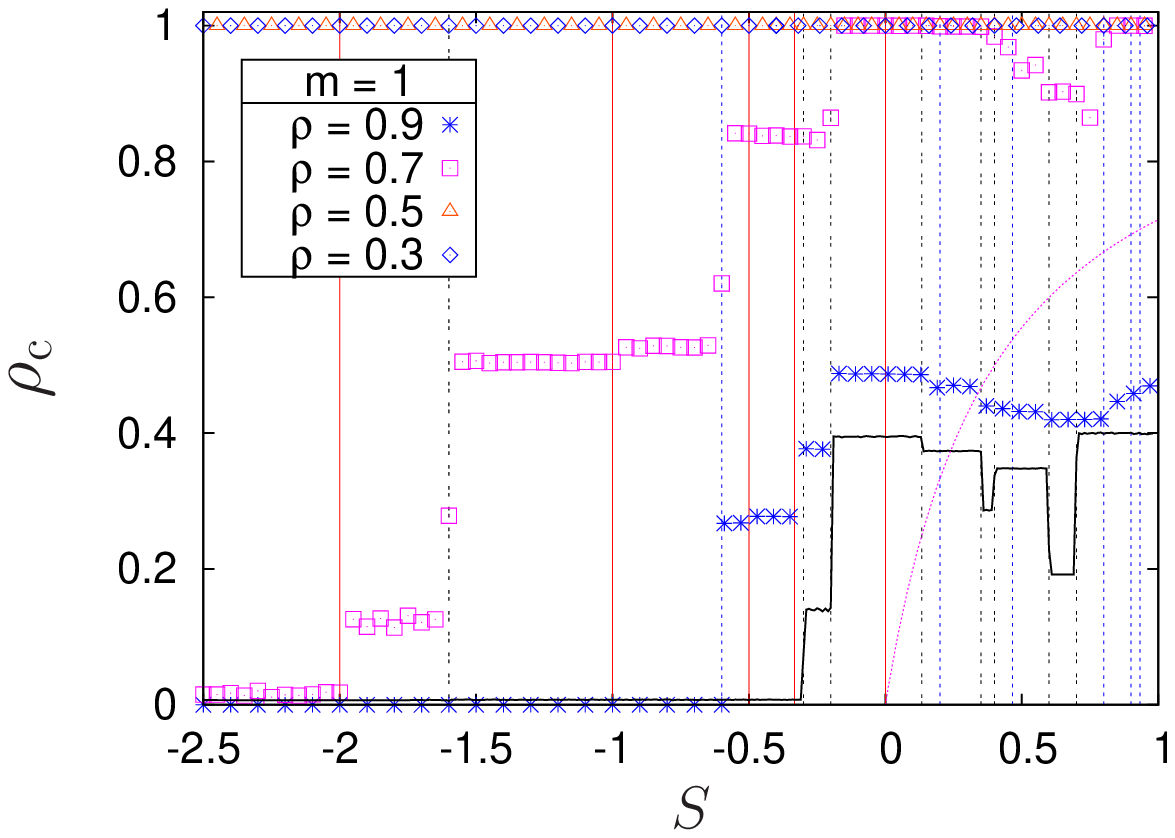}
\caption{Average fraction of cooperating individuals $\rhoc$  
versus $S$ ($T=1.4$, $R=1$ and $P=0$) for different values 
of the mobility and $\rho$ in the CDO case. The solid black line is the
case $\rho=1$, while the vertical ones locate the
transition points for both $\rho=1$ and $\rho<1$ (see fig.~\ref{fig.COD_S} and text).
Differently from the COD case, mobility introduces new transitions
(indicated by vertical solid lines) at $S=-3$, $-2$, $-1$, $-1/2$, 
$-1/3$ and 0.}
\label{fig.CDO_S}
\end{figure}

When the diffusion step is performed before the
offspring laying (CDO dynamics), the amount of cooperation is often
strongly enhanced, as can be observed in fig.~\ref{fig.CDO_S}. 
Cooperators close to defectors have low payoff; therefore, if they do not move, 
in the next step their strategy will be replaced by D. On the other hand, 
if they do move away, there is a probability of surviving depending on the 
new neighborhood they encounter (at low densities, for example, they may 
be isolated after the move, and thus survive). Moreover, 
cooperators that detach from a cooperative neighborhood have high payoff 
and may replace the defectors they find. Indeed, for many values of
the parameters $\rho$ and $m$, cooperators fully dominate
the system, even in the region of small $S$ values where the COD dynamics allows
no cooperators to survive. While all phases appearing in the COD
dynamics were already present when $m=0$, in the CDO case a few new
phases appear. These mobility driven transitions
can be seen in fig.~\ref{fig.CDO_S} and are marked by
solid vertical lines at  $S=-3$, $-2$, $-1$, $-1/2$, 
$-1/3$ and 0. In particular, a new transition appears at $S=0$ and,
differently from all other cases
where the weak PD behavior was representative of a wide range of values
of $S$, in this case, although the weak and the strict PD ($S<P$) 
still behave in the same way, the small-$S$ SD becomes different.
Even though the density of the two phases is very similar, they are
in fact different since the final configurations are slightly
different even if we prepare two systems (for example, one with
$S=-0.01$ and the other with $S=0.01$) with identical initial
conditions and subject to the same sequence of random numbers.


\section{Discussion and final comments}

The main question posed at the beginning was how mobility affects the outcome of 
different games beyond the weak dilemma at the frontier between the 
PD and the SD studied in Ref.~\cite{VaSiAr07}. A main novelty emerges
in the context of the SD game: mobility restores the enhancing
factor of the spatial structure also found in the PD game, at variance 
with the $m=0$ case where cooperation is usually lower than the fully
mixed case~\cite{HaDo04}. In general, when agents are able to
randomly diffuse on the lattice, unmatched levels of cooperation
can be attained for wide ranges of the parameters. 
Moreover, differently from the  PD and SH
games, the spatial SD presents negative responses when the value of
$S$ increases: instead of enhancing the amount of cooperation as one
would expect, $\rhoc$ sometimes decreases. This effect, absent in the 
fully mixed case, is also observed  even in the absence of mobility, 
something that has not been previously noticed. Cooperators are spatially
organized in different ways depending on the game they play. For example,
the clusters may be more compact or filamentous. This spatial structure
rules the effect that mobility has on the fate of the game.

We considered three regions of interest in the $T$ and $S$ plane, the
genuine Prisoner's Dilemma (PD) game ($T>1$ and $S<0$), the
Snowdrift (SD) game ($T>1$ and $S>0$) and the Stag Hunt (SH) game ($T<1$
and $S<0$).  
Let us analyze what happens for each of the three games
separately. We start with the genuine PD where qualitative differences 
with respect to the weak dilemma occur only for values of $S$ below a 
threshold $S^*$, a region in which cooperation is 
completely extinguished in the presence of mobility. This is reasonable since by increasing the 
penalization for the sucker's behavior (decreasing $S$) one finally
reaches a point below which C agents perform badly and cannot overcome
the filter of selection. For the COD variant at $T=1.4$, $S^* = -0.2$ no matter the 
density $\rho$ and for all $m>0$ considered. On the other hand, 
for the CDO variant, $S^*$ depends 
strongly on $\rho$. It is remarkable that, even for very severe sucker's 
penalizations (down to $S=-2.5$ in the figure, but whatever smaller value will do, and
since there is no further transition below $S=-3$, even any negativelly large one), 
for intermediate values of $\rho$ (e.g. 
$\rho = 0.5$), the universal cooperation state ($\rhoc=1$), or a state very close to that, can still be
attained.

In less severe dilemmas than the PD --- mutual defection
pays less than the sucker's payoff in the SD, and  mutual cooperation pays 
more than cheating in the SH ---  cooperation is, as one would have expected, in 
general higher. In the case of the SD, cooperation is 
often enhanced with respect to the weak dilemma with COD dynamics, while
an unprecedented state of universal cooperation ($\rhoc=1$) can be
sometimes reached with the CDO one. 
Hauert and Doebeli~\cite{HaDo04} noticed that cooperation is often inhibited
by spatial structure with $\rhoc$ being usually lower than its value in a randomly
mixed population, where for large systems one of the
three solutions $\rhoc=S/(T+S-1)$, 0 or 1 is stable.
Dilution and mobility change dramatically this scenario. When only dilution (but 
no mobility) is present, cooperation in a spatially distributed
system is higher than in the random mixed limit either for intermediate 
densities or small values of $S$.
When mobility is added, only high densities follow the behavior of the
$\rho =1$ situation where spatial structure inhibits cooperation. On the
contrary, for not so high densities cooperation is enhanced in the SD game
when $m\neq 0$.  In this way, in the
presence of mobile agents, it is again possible to make the 
statement that spatial structure promotes cooperation.
In the COD SH, the combination of mobility and large density ($\rho \ge
0.7$) leads to a boost in $\rhoc$ or even to universal cooperation. On
the other hand, for smaller values of $\rho$, provided the sucker's
payoff $S$ is also small,  $\rhoc$ is lower. So a crucial difference is 
that, for a given mobility, the level of cooperation grows with the density 
of agents, different from the behavior at $m=0$.

Accessing the actual payoff involved in real situations is not an easy task,
and it has been suggested 
that many examples that have been interpreted as
realizations of the PD are also compatible with the SD and the SH games
(see \cite{HaDo04,Skyrms04} and references therein). 
In addition to this,
mobility effects on the cooperation of real organisms are still largely unknown,
as they are difficult to isolate from other factors and, as the
theoretical results presented here have shown, even a tiny amount of
mobility is able to produce very strong changes in the final result.
Although mobility may have an effect similar to noise, 
allowing shallow basins to be avoided, they are not equivalent. 
For example, in Ref.~\cite{HaDo04},
several different dynamics, with and without noise, gave consistent results for
the inhibition of cooperation in the SD game with spatial structure, while
mobility drastically changes this outcome.
 Since the results seem to strongly depend on the chosen dynamics (although
we have only considered ``Best-takes-over'' updatings, it has two
possible variants, CDO and COD), an important,
yet open, question concerns the existence of an unifying principle, in 
Hamilton's sense~\cite{Hamilton64,Nowak06b}, relating the parameters of the game, 
that tells us when cooperative behavior might be
expected when mobile agents are present.

\begin{acknowledgments}
Research partially supported by the Brazilian agency CNPq,
grant PROSUL-490440/2007. JJA is partially supported by the 
Brazilian agencies CAPES, CNPq and FAPERGS. ES and HF want to thank
PEDECIBA for financial support.
\end{acknowledgments}


\end{document}